\documentclass[prl,twocolumn,showpacs,superscriptaddress]{revtex4-1}
\usepackage{amssymb,amsmath}
\usepackage{graphicx}
\usepackage{bm}

\newcommand{\be}{\begin{equation}}
\newcommand{\ee}{\end{equation}}
\newcommand{\mi}{\mathrm{i}}

\newcommand{\half}{\textstyle{\frac{1}{2}}}

\newcommand{\quart}{\textstyle{\frac{1}{4}}}

\begin{document}

\title{Radiative coupling and weak lasing of exciton-polariton condensates}
\author{I.~L.~Aleiner}
 \affiliation{Physics Department, Columbia University, New York, NY 10027, USA}
\author{B.~L.~Altshuler}
 \affiliation{Physics Department, Columbia University, New York, NY 10027, USA}
\author{Y.~G.~Rubo}
 \affiliation{Centro de Investigaci\'on en Energ\'{\i}a, Universidad Nacional
 Aut\'onoma de M\'exico, Temixco, Morelos, 62580, Mexico}


\begin{abstract}
In spite of having finite life-time exciton-polaritons in microcavities
are known to condense at strong enough pumping of the reservoir. We
present an analytical theory of such Bose-condensates on a set of
localized one-particle states: condensation centers. To understand
physics of these arrays one has to supplement the Josephson coupling by
the radiative coupling caused by the interference of the light emitted
by different centers. Combination of these couplings with the one-site
interaction between the bosons leads to a rich nonlinear dynamics. In
particular, a new regime of radiation appears. We call it weak lasing:
the centers have macroscopic occupations and radiate coherently, but
the coupling alone is sufficient for stabilization. The system can have
several stable states and switch between them. Moreover, the time
reversal symmetry in this regime is, as a rule, broken.
\end{abstract}

\maketitle

\emph{Introduction.}---Condensation of exciton-polaritons (EP) in semiconductor
microcavities formed by two Bragg mirrors with a quantum well between them was
recently discovered \cite{Kasprzak06, Balili07, Lai07, Baumberg08, Wertz09}. As
experimental implementation of the Bose-Einstein condensation (BEC) these
systems have some advantages as compared with cold gases, e.g., the vortex
dynamics and the superfluid motion can be accessed by optical methods
\cite{Lagoudakis0809, Roumpos11, Amo09}. At the same time, in contrast with the
atomic BEC the EP-condensates are not in the thermodynamic equilibrium. To
reach a macroscopic number of EPs, which life-time is finite, one needs an
outside pumping. Being driven, the EP-condensates differ fundamentally from the
conventional BEC-systems. In particular, EPs can condense into one-particle
excited state \cite{Lai07, Manni11} or even into several excited states
\cite{Love08, Krizh09}.

It is not unusual for a one-particle bosonic state to be localized.
Such states either formed by disorder or intentionally prepared
\cite{Balili07, Lai07} can serve as condensation centers (CC). The
bosons (photons, excitons, polaritons, etc.) arrive to each CC from an
incoherent reservoir created by the pumping and escape in the form of
light radiation. At low enough pumping one should expect a system of
disconnected BEC droplets emitting light of different frequencies. As
the pumping increases these sources of radiation tend to become
coherent and to shine as a laser. This effect resembles the
Insulator-Superfluid transition, but the similarity is limited by the
dissipative nature of the system. In this paper we develop the theory
of lasing in non-equilibrium BEC, which allows one to understand
qualitatively and describe quantitatively a number of unusual and
surprising experimental observations.

Existing theoretical approach to non-equilibrium EP condensates
\cite{Wouters07,Keeling08} is based on the Gross-Pitaevskii equation (GPE)
modified to account for the finite life-time of the bosons, the continuous feed
of the condensate from the reservoir, and effects of reservoir depletion.
Numerical analysis seems to agree qualitatively with some experiments. However,
the nonlinear GPE can have several distinct solutions and the choice is
ambiguous.

Our starting idea is that the sensitivity of the bosonic life-time to the
symmetry of the wave-function dominates this choice. If two CCs are close to
the resonance and not too far apart, the interference of the light emitted from
different CCs is constructive for symmetric (bonding) state and destructive for
antisymmetric (anti-bonding) state, i.e., the bosons live longer in the
anti-bonding case. Existence of several states with different life-times leads
to a crossover range of pumping strengths (rather than a single threshold)
where the income and outcome rates match. Within this range a coherent
condensed state is formed. In this respect, the non-equilibrium EP condensation
in disordered cavities resembles the random lasing phenomena: the lasing states
also possess the longest life-times. Important difference is the nonlinearity:
interaction between the EPs is able to synchronize the frequencies of CCs
\cite{WouterSyn,EasthamSyn}, which have no reason to be in resonance at low
occupancies. One can call this regime ``weak lasing'': the occupations of the
CCs are macroscopic and the radiation is coherent, but the state is stabilized
by the coupling between CCs rather than, e.g., by the depletion of the
reservoir.

Below we show that CCs indeed become coherent and form a particular long-living
condensate. In certain range of the parameters the system can have several
linearly stable states, and switch between them. We argue that the properties
of these states and the switches naturally explain otherwise mysterious
experimental observations.

\emph{Formalism.}---In the familiar Glauber-Sudarshan formalism of coherent
states \cite{MandelWolfBook} the occupation $n$ and the phase $\phi$ of each CC
are encoded in the complex number $z=\sqrt{n}\,e^{\mi\phi}$. $N$ CCs are thus
characterized by $N$-component complex vector
$\mathbf{Z}=\{z_1,z_2,\dots,z_N\}$, where $z_\mu$ corresponds to the CC number
$\mu$. Each state of this system is fully described by the density matrix
$\rho(\mathbf{Z},\mathbf{Z}^*,t)$, which evolves in time according to
($\hbar=1$)
\begin{equation}
 \label{RhoEqGen}
 \dot{\rho}=\sum_\mu\left[ W_\mu\partial_\mu^*\partial_\mu\rho
 +2\Re\left\{
 \partial_\mu\!\left(\rho\partial_\mu^*\mathcal{H}\right)
 \right\}\right].
\end{equation}
Here $W_\mu$ is the incoming rate of the bosons to the $\mu$-th CC,
$\partial_\mu=\partial/\partial z_\mu$, dot indicates the time derivative, and
$\mathcal{H}$ is the complex Hamiltonian function
\begin{equation}
 \label{ComplexHN}
 \mathcal{H}(\mathbf{Z},\mathbf{Z}^*)
 =\sum_\mu\mathcal{H}^{(1)}(|z_\mu|^2)+\frac{1}{2}\sum_{\mu\ne\nu}V_{\mu\nu}z_\mu^*z_\nu.
\end{equation}
The Hamiltonian function of an isolated CC is
\begin{equation}
 \label{ComplexH1}
 \mathcal{H}^{(1)}(|z_\mu|^2)=\half\left(\Gamma_\mu-W_\mu\right)|z_\mu|^2
 +\mi H_\mu(|z_\mu|^2),
\end{equation}
where $\Gamma_\mu$ is the escape rate, $H_\mu(n_\mu)$ is the energy of bosons
in the $\mu$-th CC, which alone emits light with the frequency
$\Omega_\mu=dH_\mu(n_\mu)/dn_\mu$. For weak interaction
\begin{equation}
 \label{HEnergy}
 H_\mu(n_\mu)=\omega_\mu n_\mu+\quart\alpha_\mu n_\mu^2,
 \qquad
 \Omega_\mu=\omega_\mu+\half\alpha_\mu n_\mu,
\end{equation}
with one-particle energy $\omega_\mu$ and the coupling constant $\alpha_\mu>0$.
The coupling is weak, $\alpha_\mu\ll\omega_\mu$, and $\Omega_\mu$ slightly
increases with the pumping. In general, the income rates $W_\mu$ are dynamical
variables determined by the occupations of the reservoir states. However, in
(1-3) $W_\mu$ enter as $\mathbf{Z}$-independent external parameters. We adopt
this simplifying assumption, which is valid when the life-time of the reservoir
excitations is not limited by the inelastic relaxation into CCs. It is quite
straightforward to generalize our approach to include the dynamics of $W_\mu$.

The coupling $V_{\mu\nu}$ between CCs in \eqref{ComplexHN} is bilinear, i.e.,
the bosons from different CCs do not interact, and it consists of radiative
(dissipative) and Josephson (non-dissipative) parts,
$V_{\mu\nu}=\gamma_{\mu\nu}-\mathrm{i}J_{\mu\nu}$. The Josephson coupling
favors symmetric ground state, i.e., $J\geqslant0$. The radiative coupling
$\gamma_{\mu\nu}$ is due to the interference of the light from different CCs.
The matrices $\gamma_{\mu\nu}$ and $J_{\mu\nu}$ are Hermitian and have zero
diagonal matrix elements, which are absorbed by the first term in
\eqref{ComplexHN}. Besides, we assume no explicit time reversal symmetry
violation and thus $\gamma_{\mu\nu}$ and $J_{\mu\nu}$ are real. Note that the
Josephson coupling caused by the tunneling is exponentially small, when CCs are
spatially separated, while $\gamma_{\mu\nu}$ can be still substantial.

The time evolution of the density matrix without Josephson coupling should
reflect the following physical picture. According to \eqref{ComplexHN} and
\eqref{ComplexH1} both $\Re z_\mu$ and $\Im z_\mu$ oscillate with the high
frequency $\Omega_\mu$. As long as $\Omega_\mu\ne\Omega_\nu$, coupling between
the CCs $\mu$ and $\nu$ averages out as time exceeds
$|\Omega_\mu-\Omega_\nu|^{-1}$. However the interaction between the bosons
synchronizes the frequencies. As soon as this happens, the interference
suppresses the escape if the phases of CCs are opposite. This suppression in
turn stabilizes this nontrivial stationary state (NTSS) with high occupations
and coherent radiation of the CCs. NTSS differs drastically from the trivial
stationary state (TSS) with small $n_\mu$ and different emission frequencies.
We emphasize that NTSSs appear due to the off-diagonal radiative coupling even
at $J_{\mu\nu}=0$.

At this point we adopt the Langevin approach, which remains valid for a complex
Hamiltonian function. We start with the formal solution of Eq.~\eqref{RhoEqGen}
for $\rho(\mathbf{Z},\mathbf{Z}^*,t)$ with the initial condition
$\rho_0(\mathbf{Z},\mathbf{Z}^*)$ in the form of Feynman path integral over the
``trajectories'' $\mathbf{Z}(t)$
\begin{equation}
 \label{FeynmanPI}
 \rho(\mathbf{Z},\mathbf{Z}^*,t)=
 \int\!\rho_0(\mathbf{Z}_0,\mathbf{Z}_0^*)d^{2N}\!\mathbf{Z}_0\!\!
 \int_{\mathbf{Z}(0)=\mathbf{Z}_0}^{\mathbf{Z}(t)=\mathbf{Z}}
 \mathcal{D}\mathbf{Z}(\tau) e^{-\mathbb{S}},
\end{equation}
\vspace{-0.5\baselineskip}
\begin{equation}
\label{Action}
 \mathbb{S}=\sum_\mu\int_{0}^t
 \frac{1}{W_\mu}\left|
 \dot{z}_\mu(\tau)+\partial_\mu^*\mathcal{H}
 \right|^2 d\tau,
\end{equation}
Note that the pumping intensity $W_\mu$ of the  $\mu$-th CC plays the role of
its effective temperature.

One can rewrite \eqref{FeynmanPI} as the average
$\left<\delta(\mathbf{Z}-\mathbf{Z}^{(f)}(t))\right>$, where $z_\mu^{(f)}(t)$
are the solutions of equations
\begin{equation}
 \label{Langevin}
 \dot{z}_\mu+\partial_\mu^*\mathcal{H}
 =f_\mu(t),
\end{equation}
with $f_\mu(t)$ being a realization of the Gaussian random processes
with zero mean $\left<f_\mu(t)\right>=0$ and $\delta$-like two-point
correlation function
$\left<f_\mu(t)f_\nu^*(t^\prime)\right>=W_\mu\delta_{\mu\nu}\delta(t-t^\prime)$.

Our analysis of \eqref{Langevin} consists of the following steps. First we omit
noise to find the stationary points. Next we analyze the stability of these
points and thus identify the long-living stationary states. If such states
coexist, the time the system spends in each basin of attraction is finite due
to the noise $f_\mu(t)$. The set of these times determines the behavior of the
system. Below we evaluate these times for the simplest case. The final
step---the analysis of the fluctuations of $\mathbf{Z}$ in the vicinity of the
stationary points, which determine the line-shapes of the radiation and its
statistics---lies beyond the scope of this letter and will be discussed
elsewhere.

Substitution of (2,3) into \eqref{Langevin} at $f_\mu=0$ leads to
\begin{equation}
 \label{CoupledGPEs}
 \dot{z}_\mu=-(\Gamma_\mu-W_\mu+2\mi\Omega_\mu)\frac{z_\mu}{2}
 -\sum_{\mu\ne\nu}(\gamma_{\mu\nu}-\mi J_{\mu\nu})\frac{z_\nu}{2}.
\end{equation}
Being similar to the equations used in \cite{WouterSyn} Eqs.\
\eqref{CoupledGPEs} differ in two aspects. First one is the radiative coupling.
The second difference is the absence of the reservoir dynamics. We believe that
the depletion of reservoir is not qualitatively important in the weak lasing
regime.

\emph{Two Condensation Centers.}---From now on we restrict ourselves by the
case of two coupled CCs. The density matrix $\rho^{(2)}(z_1,z_2)$ depends on
four real variables, however, the total phase, is irrelevant to what follows.
We choose to parameterize the remaining degrees of freedom by three components
of a 3D classical spin $\mathbf{S}=(S_x,S_y,S_z)$ using the Pauli matrices
$\bm{\sigma}_{\mu\nu}$,
\begin{equation}
 \label{Spin}
 \mathbf{S}=\frac{1}{2}\sum_{\mu,\nu=1,2} z_\mu^*\bm{\sigma}_{\mu\nu}z_\nu,
\end{equation}
with $S^2=S_x^2+S_y^2+S_z^2=(|z_1|^2+|z_2|^2)^2/4$. For simplicity we assume
that $(\Gamma_1-W_1)=(\Gamma_2-W_2)=g$; a small difference between
$(\Gamma_1-W_1)$ and $(\Gamma_2-W_2)$ would not affect the results
qualitatively. From \eqref{CoupledGPEs} and \eqref{Spin} we have
\begin{equation}
 \label{SpinDyn}
 \dot{\mathbf{S}}=-S\bm{\nabla}(\Re\mathcal{H})
 -\left[\mathbf{S}\times\bm{\nabla}(\Im\mathcal{H})\right],
\end{equation}
where $\nabla=\partial/\partial\mathbf{S}$. Calling $\omega=\omega_1-\omega_2$,
$\alpha=(\alpha_1+\alpha_2)/2$, and $\beta=(\alpha_2-\alpha_1)/2$, we rewrite
\eqref{SpinDyn} explicitly
\begin{subequations}
 \label{SpinDynComp}
 \begin{eqnarray}
 &&\dot{S}_x=-g S_x-\gamma S-\left[\omega+\alpha S_z-\beta S\right]S_y, \\
 &&\dot{S}_y=-g S_y+JS_z+\left[\omega+\alpha S_z-\beta S\right]S_x, \\
 &&\dot{S}_z=-g S_z-JS_y,
 \end{eqnarray}
\end{subequations}
and accordingly $\dot{S}=-g S-\gamma S_x$. We used
$\gamma_{\mu\nu}=\gamma\sigma^x_{\mu\nu}$ and $J_{\mu\nu}=J\sigma^x_{\mu\nu}$.

The stationary states are solutions of (\ref{SpinDynComp}) at
$\dot{\mathbf{S}}=0$. It turns out that, apart from TSS $\mathbf{S}=0$, as long
as $0<g<\gamma$ there exist NTSSs. In polar coordinates $(S,\vartheta,\varphi)$
their positions are
\begin{subequations}
 \label{NTSSs}
\begin{eqnarray}
 \varphi=\pi-\arctan r, \quad \cos\vartheta=-Jr/\gamma, \\
 S=\frac{\gamma[g\omega+(J^2+g^2)r]}{g[\beta\gamma+\alpha Jr]},
 \quad
 r^2=\frac{\gamma^2-g^2}{J^2+g^2}.
\end{eqnarray}
\end{subequations}

Since $r=\pm|r|$, Eq.\ (\ref{NTSSs}b) for $S$ gives two solutions. Each
\emph{positive} solution corresponds to a NTSS. $S$ changes the sign together
with either its numerator or its denominator, i.e., $g$ has two bifurcation
values, $g_0$ and $g_\infty$:
\begin{subequations}
 \label{Bifurcations}
 \begin{eqnarray}
 \omega^2g_0^2=(J^2+g_0^2)(\gamma^2-g_0^2), \\
(\beta^2\gamma^2+\alpha^2 J^2)g_\infty^2=\gamma^2 J^2(\alpha^2-\beta^2).
\end{eqnarray}
\end{subequations}
At $g=g_0$ one of the NTSSs merges with the TSS. At $g=g_\infty$ and $r<0$ the
value of $S$ diverges.

To manifest itself as an attractor the NTSS has to be linearly stable. The
stability diagram Fig.\ 1 follows from the analysis of the Lyapunov exponents.
The TSS loses its stability at $g_0$. The stability of NTSS with a given $S$
(\ref{NTSSs}b) changes at $g=g_c$ defined by the equation
\begin{equation}
 \label{Stability}
 \left[(J^2-g^2)\beta\gamma r+(\gamma^2+g^2)\alpha J\right]gS
 =2\gamma(g^4+\gamma^2J^2).
\end{equation}

\begin{figure}[t]
\includegraphics[width=3.2in]{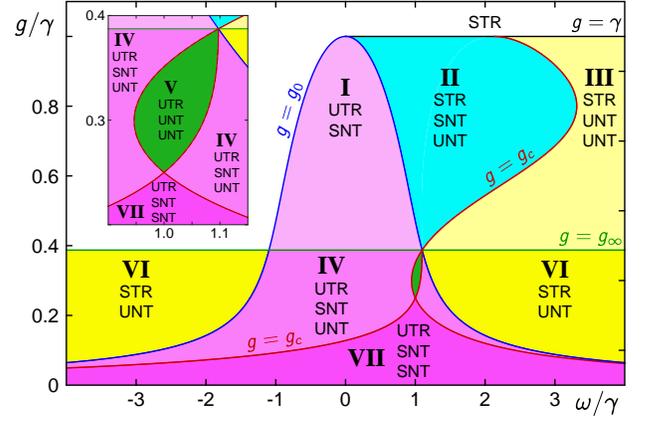}    
 \caption{Bifurcation and stability curves, according to Eqs.\ \eqref{Bifurcations}
 and \eqref{Stability}, for $J/\gamma=0.25$ and $\alpha=2\beta$.
 Stable(unstable) TSS and NTSS are indicated by STR(UTR) and SNT(UNT) in each domain.
 Zoomed domain V is shown in the insert.}
\end{figure}

Here we are not aiming to describe the nonlinear dynamics of the two CCs system
in all details. In particular, we discussed only the stationary states.
However, there are numerical evidences for the existence of the limiting cycle
(LC) in the regions IV and V in Fig.\ 1. Moreover, within the region V all of
the stationary states are unstable, i.e., there should be at least one stable
LC. Analysis of stable \emph{time-dependent} solutions and their experimental
manifestations is a subject for further studies.

\emph{Kramers transitions.}---In the regions II and VII in Fig.1 two stable
stationary states coexist, which leads to an additional problem: the noise term
in \eqref{Langevin} causes switching  between these states. The properties of
the radiation from the two states are very different. If both escape rates
$\tau^{-1}$ are small enough, this switching can be observed directly.
Otherwise we expect a combination of the two signals, the weights being
determined by the relation between the escape rates.

One can evaluate $\tau^{-1}$ analytically when $J=0$ and according to
(\ref{Bifurcations}b) $g_\infty=0$. In this case the region III disappears and
the bistable region II is defined by the inequalities $\omega>\gamma>g>0$. At
$g=0$ the system is conservative and integrable, i.e., (\ref{SpinDynComp}) has
two integrals of motion: $S_z$ [see (\ref{SpinDynComp}c)] and $A=2\beta(\gamma
S_y+\omega S)-\beta^2 S^2$. Stationary states correspond to $S_z=0$ for all
$g>0$. For TSS $A=0$, while for NTSSs $A=(\omega\pm\gamma)^2=A_\pm$ (the stable
and unstable NTSS correspond to $+$ and $-$, respectively). Other values of $A$
at $g=0$ label periodic trajectories (see Fig.\ 2). Slow time evolution of $A$
at small $g>0$ is governed by
\begin{equation}
 \label{EqForA}
 \dot{A}=\frac{2gE}{K_{-1}}+F(t), \quad
 \left<F(t)F(t^\prime)\right>=\beta T\delta(t-t^\prime),
\end{equation}
\vspace{-\baselineskip}
\begin{equation}
 \label{ANoise}
 T=\frac{4W}{K_{-1}}\left[
 2\omega(E+K_{-1}A)+(\gamma^2-\omega^2-A)\frac{\partial E}{\partial\omega}
 \right],
\end{equation}
where $E=K_{+1}-2\pi\omega\theta(A_--A)$,
\begin{equation}
 \label{KIntegrals}
 K_{\pm1}=\int_0^{2\pi}\!\!R^{\pm1/2}\theta(R)d\varphi, \quad
 R=(\omega+\gamma\sin\varphi)^2-A,
\end{equation}
with $\theta(x>0)=1$ and $\theta(x<0)=0$.

Equation \eqref{EqForA} is a typical case of the one-dimensional Kramers
problem---classical transitions between two potential minima due to the thermal
noise---slightly modified by the fact that ``effective temperature''
\eqref{ANoise} depends on the ``coordinate'' $A$. Using the familiar solution
\cite{Langer69,Dykman84,Hanggi90} we estimate the escape rates from the stable
TSS and NTSS as
\begin{equation}
 \label{EscapeRates}
 \frac{\beta}{W\omega}
 \exp\left\{-\frac{2g}{\beta}\int_{A_u}^{A_s}\frac{EdA}{K_{-1}T}\right\},
\end{equation}
where $A_u=A_-=(\omega-\gamma)^2$, $A_s=0$ and $A_s=A_+=(\omega+\gamma)^2$ for
TSS and NTSS, respectively. This approach is valid only provided that the
exponential factor is small, i.e., when $\omega/\beta\gg1$ and thus the
occupations at NTSS are large.

One can use \eqref{EscapeRates} to compare the steady state occupations of TSS
and NTSS. It turns out that for small $\omega$ the NTSS prevails. The escape
rate from NTSS is approximately independent of $\omega$, while the escape rate
from TSS decreases with $\omega$. The rates become equal at
$\omega\simeq4.1\gamma$. For bigger $\omega$ TSS dominates.

\begin{figure}[t]
\includegraphics[width=3.3in]{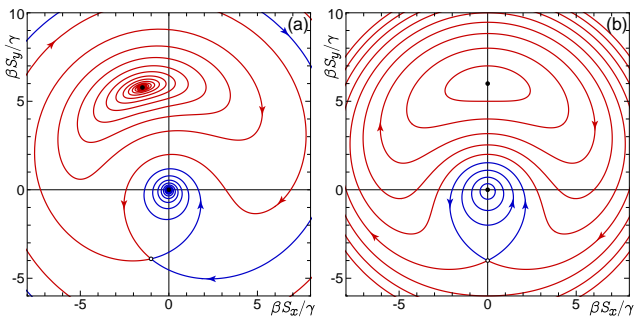}    
 \caption{Flow near NTSSs and TSS for $J=0$ and $\omega=5\gamma$.
 (a) The saddle trajectories for $g=0.25\gamma$.
 (b) Several closed trajectories for $g=0$.}
\end{figure}

\emph{Comparison to experimental observations.}---We believe that a number of
recently observed features of the radiation of the disordered exciton-polariton
structures can be naturally interpreted in the framework of our theory. Large,
$\pi/2<\varphi<3\pi/2$, phase differences between the CCs, which manifests
itself in the dip of the radiation intensity $I(\mathbf{k}_\perp)$ at small
transverse wave-numbers $\mathbf{k}_\perp$ was reported by several groups
\cite{Lai07,Manni11,Love08,Krizh09}. A detailed analysis of the radiation from
disordered CdTe structures was performed in \cite{Krizh09}. It was found that
(i) the number of CCs differs from the number of radiated frequencies; (ii)
some frequencies are radiated from several CCs, and (iii) its distribution in
the $\mathbf{k}_\perp$-space as a rule has a characteristic annular shape. The
last but not the least important observation in \cite{Krizh09} is (iv) the
clear absence of the $\mathbf{k}_\perp\rightarrow -\mathbf{k}_\perp$ symmetry,
$I(\mathbf{k}_\perp)\ne I(-\mathbf{k}_\perp)$.

Features (i), (ii) is exactly what we should expect: for example, as long as
TSS and NTSS in the system of two CCs coexist each CC radiates its own
frequency and in addition they together radiate the third one, i.e. two CCs
radiate three frequencies and one of them is radiated from both of them. The
number of collectively radiated frequencies quickly increases with the number
of CCs. (iii) had already been discussed -- it reflects the large phase
differences favored by the radiative coupling.

(iv) The time inversion symmetry implies that the phase differences between
different CCs can be either 0 or $\pi$. One can check that under this condition
$I(\mathbf{k}_\perp)=I(-\mathbf{k}_\perp)$. However, according to
(\ref{NTSSs}a) in the NTSS the phase mismatch $\varphi=\phi_2-\phi_1$ between
the two CCs is neither 0 nor $\pi$. We believe that by observing states with
broken time inversion symmetry authors of \cite{Krizh09} experimentally proved
existence of the radiative coupling.

\emph{In conclusion}, we considered radiation of coupled centers of
condensation of exciton-polaritons in the weak lasing regime, where the
occupations $n_1$ and $n_2$ of the centers are macroscopically large, while the
occupations of the states of the incoherent reservoir remain independent on
$n_{1,2}$. Apart from the usual Josephson coupling between the centers we took
into account the radiative coupling due to the interference of the two sources
of light. The radiative coupling turns out to be responsible for the very
existence of the weak lasing regime and crucial for understanding the
experiments on coupled condensates. The onsite interaction between the bosons
makes the dynamics of the system nonlinear and rich. In particular, depending
on the parameters there can be more than one stable stationary states, and some
of them are nontrivial: in contrast with the trivial stationary state
(independent sources of radiation) the radiation of the two centers is
synchronized in frequency. The phase difference is locked thus breaking the
time-reversal symmetry. This symmetry breaking was recently observed
\cite{Krizh09}.

We acknowledge the discussions with S.\ Flach, A. V. Kavokin, V. E. Kravtsov,
D. N. Krizhanovskii, A.\ Pikovsky, G. V. Shlyapnikov, and M.\ S.\ Skolnick.
This work was supported in part by NSF-CCF Award 1017244 and by DGAPA-UNAM
under the Grant IN112310.


\end{document}